\documentclass[journal=jcisd8]{achemso}
\usepackage[pdftex,colorlinks=true,linkcolor=black,citecolor=black,urlcolor=black,bookmarks=true,breaklinks=true]{hyperref}
\usepackage[usenames,pdftex]{color}
\usepackage{amsthm,amsmath,amssymb,multirow}
\usepackage{xcolor,colortbl}
\usepackage{subcaption}
\usepackage{tikz}
\setkeys{acs}{maxauthors=0}
\SectionNumbersOn

\title{Quantitative toxicity prediction using topology based multi-task deep neural networks}

\author{Kedi Wu}
\affiliation{Department of Mathematics, Michigan State University, MI 48824, USA}
\author{Guo-Wei Wei}
\affiliation{Department of Mathematics, Michigan State University, MI 48824, USA}
\alsoaffiliation{Department of Electrical and Computer Engineering, Michigan State University, MI 48824, USA}
\alsoaffiliation{Department of Biochemistry and Molecular Biology, Michigan State University, MI 48824, USA}
\email{wei@math.msu.edu}

\begin{document}

\begin{abstract}
The understanding of toxicity is of paramount importance to  human health and environmental protection. Quantitative toxicity analysis has become a new standard in the field.   This work introduces element specific persistent homology (ESPH), an algebraic topology approach, for quantitative toxicity prediction. ESPH retains crucial chemical information during the topological abstraction of geometric complexity and provides a representation of small molecules that cannot be obtained by any other method. To investigate the representability and  predictive power of ESPH for small molecules, ancillary   descriptors have also been developed based on physical models. Topological and physical descriptors are paired with advanced machine learning algorithms, such as  deep neural network (DNN),  random forest (RF) and gradient boosting decision tree (GBDT), to facilitate their applications to quantitative toxicity predictions. A topology based multi-task strategy is proposed to take the advantage of the availability of large data sets while dealing with small data sets. Four benchmark toxicity data sets that involve quantitative measurements are used  to validate  the proposed approaches. Extensive numerical studies indicate that the proposed topological learning methods are able to outperform the state-of-the-art methods in the literature for quantitative toxicity analysis.  Our online server for computing element-specific topological descriptors (ESTDs) is available at {\url{http://weilab.math.msu.edu/TopTox/}}.

\end{abstract} 

\maketitle 

\section{Introduction}\label{sec:Intro}

Toxicity is a measure of the degree to which a chemical can adversely affect an organism. These adverse effects, which are called toxicity endpoints, can be either quantitatively or qualitatively measured by their effects on given targets.  
Qualitative toxicity  classifies  chemicals into toxic and nontoxic categories, while  quantitative toxicity data set records the minimal amount of  chemicals  that can reach certain lethal effects. Most toxicity tests aim to  protect human from harmful effects caused by chemical substances and are traditionally conducted in {\sl in vivo} or {\sl in vitro} manner. Nevertheless, such experiments are usually very time consuming and cost intensive, and even give rise to ethical concerns when it comes to animal tests. Therefore, computer-aided methods, or {\sl in silico} methods, have been developed to improve prediction efficiency without sacrificing too much of accuracy. Quantitative structure activity relationship (QSAR) approach is one of the most popular and commonly used approaches. The basic QASR assumption is that similar molecules have similar activities. Therefore by studying the relationship between chemical structures and biological activities, it is possible to predict the activities of new molecules without actually conducting lab experiments.  

There are several types of algorithms to generate QSAR models: linear models based on linear regression and linear discriminant analysis \cite{deeb:2012};  nonlinear models including nearest neighbor \cite{kauffman:2001, ajmani:2006}, support vector machine  \cite{deeb:2012,si:2007, du:2008} and random forest \cite{svetnik:2003}. These methods have advantages and disadvantages \cite{liupx:2009} due to their statistics natures. For instance, linear models overlook the relatedness between different features, while nearest neighbor method largely depends on the choice of descriptors. To overcome these difficulties, more refined and advanced machine learning methods have been introduced. Multi-task (MT) learning   \cite{caruana:1998} was proposed partially to deal with data sparsity problem, which is commonly encountered in QSAR applications. The idea of MT learning is to learn the so-called ''inductive bias" from related tasks to improve accuracy using the same representation. In other words, MT learning aims at learning a shared and generalized feature representation from multiple tasks. Indeed, MT learning strategies have brought new insights to bioinformatics since compounds from related assays may share features at various feature levels, which is extremely helpful if data set is small. Successful applications include splice-site and MHC-I binding prediction \cite{widmer:2012} in sequence biology, gene expression analysis, and system biology \cite{xuq:2011}. 

Recently, deep learning (DL) \cite{lecun:2015,schmidhuber:2015}, particularly convolutional neural network (CNN), has emerged as a powerful paradigm to render a wide range of the-state-of-the-art results in signal and information processing fields, such as speech recognition \cite{dahl:2012, deng:2013} and natural language processing  \cite{socher:2012, sutskever:2014}. Deep learning architecture is essentially based on artificial neural networks. The major difference between deep neural network (DNN)  models and non-DNN models is that DNN models consist of a large number of layers and neurons, making it possible to construct abstract features.   

Geometric representation of molecules often contains too much structural detail  and  thus is prohibitively expensive for most realistic large molecular systems. However, traditional topological methods often reduce too much of   the original geometric  information. Persistent homology, a relatively new branch of algebraic topology,  offers an  interplay between  geometry and topology \cite{Edelsbrunner:2002, zomorodian:2005}. It creates a variety of topologies of a given object by varying a  filtration parameter. As a result, persistent homology can capture topological structures continuously over a range of spatial  scales. Unlike commonly used computational homology which results in  truly metric free  representations, persistent homology embeds geometric information in topological invariants, e.g., Betti numbers,  so that ``birth"  and ``death" of  isolated components,  rings,  and cavities can be monitored at all geometric scales by topological measurements. 

Recently,  we have introduced persistent homology  for the modeling and characterization of  nano particles, proteins and other biomolecules \cite{KLXia:2014c, KLXia:2015a, KLXia:2015d,KLXia:2015e,KLXia:2015b, BaoWang:2016a}. We proposed  molecular topological fingerprint (TF)   to reveal topology-function relationships in protein folding and protein flexibility \cite{KLXia:2014c}.  This approach was integrated  machine-learning algorithms for protein classification \cite{ZXCang:2015}. However, it was found that  primitive persistent homology has a limited power in protein classification due to its oversimplification of biological information \cite{ZXCang:2015}. Most recently, element specific persistent homology (ESPH) has been introduced to retain crucial biological information during the topological simplification of geometric complexity   \cite{ZXCang:2017a, ZXCang:2017b, ZXCang:2017c}. The integration of ESPH and machine learning gives rise to some of the most accurate predictions of protein-ligand binding affinities  \cite{ZXCang:2017b, ZXCang:2017c} and mutation induced protein stability changes  \cite{ZXCang:2017a,  ZXCang:2017c}. However, ESPH has not been validated for its potential utility in small molecular characterization, analysis, and modeling. In fact, unlike proteins, small molecules involve a large number of element types and are more diversified in their chemical compositions. They are also rich in structural  variability in structures, including cis-trans distinctions and chiral and achiral stereoisomers. Small molecular properties are very sensitive to their structural and compositional differences. Therefore, it is important to understand the representability and predictive power of ESPH in dealing with small molecular  diversity, variability and sensitivity.   

The  objective for this work is to introduce   element specific topological descriptors (ESTDs) constructed via ESPH  for  quantitative toxicity analysis and prediction of small molecules.  We explore the representational and predictive powers of  ESTDs for small molecules. Physical descriptors constructed from microscopic models are also developed both as ancillary descriptors and as competitive descriptors to further investigate the proposed topological methods. These new descriptors are paired with advanced machine learning algorithms,  including MT-DNN, single-task DNN (ST-DNN), random forest (RF) and  gradient boosting decision tree (GBDT), to construct topological learning strategies for   illustrating their predictive power in quantitative toxicity analysis. We demonstrate that the proposed topological learning provides a very competitive description of relatively small drug-like molecules. Additionally, the inherent correlation among different quantitative toxicity endpoints makes our topology based  multitask strategy a viable approach to quantitative toxicity  predictions.

\section{Methods and algorithms} \label{sec:methods}

In this section,  we provide a detail discussion about  molecular descriptors used in this study, including element-specific topological descriptors and auxiliary descriptors calculated from physical models.   Moreover,  an overview of machine learning algorithms, including ensemble methods (random forest and gradient boosting decision tree),  deep neural networks, single-task learning and multi-task learning, is provided. Emphasis is given to  advantages of multi-task deep convolutional neural network for quantitative toxicity endpoint predictions and how to select appropriate parameters for network architectures. Finally, we provide a detailed description of our learning architecture, training procedure and evaluation criteria.

\subsection{Element specific topological descriptor (ESTD)} \label{topo_feature}
In this subsection, we give a brief introduction to persistent homology and  ESTD construction. An example is also given to illustrate the   construction.

\subsubsection{Persistent homology}

For atomic coordinates in  a molecule, algebraic groups can be defined via simplicial complexes, which are constructed from simplices, i.e., generalizations of the geometric notion of nodes, edges, triangles, tetrahedrons, etc. Homology associates a sequence of algebraic objects, such as abelian groups, to topological spaces and characterizes the topological connectivity of geometric objects in terms of topological invariants, i.e., Betti numbers, which are used to distinguish topological spaces.  Betti-0, Betti-1 and Betti-2, respectively, represent independent components, rings and cavities in a physical sense. A filtration parameter, such as the radius of a ball, is used to continuously vary over an interval so as to generate a family of structures. Loosely speaking, the corresponding family of homology groups induced by the filtration is a persistent homology. 
The variation of the topological invariants, i.e.,  Betti numbers, over the filtration gives rise to a unique characterization of physical objects, such as molecules.

\paragraph{Simplex} Let $u_0, u_1, \ldots, u_k$ be a set of points in $\mathbb{R}^d$. A point $x=\sum_{i=0}^k \lambda_iu_i$ is called an {\it affine combination} of the $u_i$ if $\sum_{i=0}^k\lambda_i=1$. The $k+1$ points are said to be {\it affinely independent}, if and only if $u_i-u_0$, $1 \le i \le k$ are linearly independent. We can find at most $d$ linearly independent vectors and at most $d+1$ affinely independent points in $\mathbb{R}^d$. \par
\begin{figure}[ht!]
\small
\centering
\begin{subfigure}{0.24\textwidth}
\centering
\includegraphics[scale=2]{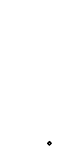}
\caption{0-simplex}
\label{fig:dot}
\end{subfigure}
\centering
\begin{subfigure}{0.24\textwidth}
\centering
\includegraphics[scale=2]{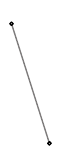}
\caption{1-simplex}
\label{fig:line}
\end{subfigure}
\centering
\begin{subfigure}{0.24\textwidth}
\centering
\includegraphics[scale=2]{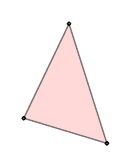}
\caption{2-simplex}
\label{fig:triangle}
\end{subfigure}
\centering
\begin{subfigure}{0.2\textwidth}
\centering
\includegraphics[scale=2]{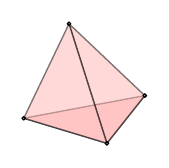}
\caption{3-simplex}
\label{fig:tet}
\end{subfigure}
\caption{Examples of simplex of different dimensions.  (a), (b), (c) and (d) above represent  0-simplex, 1-simplex, 2-simplex, and 3-simplex, respectively. }
\label{fig:ph}
\end{figure}

 
An affine combination, $x=\sum_{i=0}^k \lambda_iu_i$ is a {\it convex combination} if $\lambda_i$ are nonnegative. A {\it $k$-simplex}, which is defined to be the {\it convex hull} (the set of convex combinations) of $k+1$ affinely independent points,  can be formally represented as
\begin{equation}\label{simplex}
\sigma=\left\{\sum_{i=0}^k\lambda_iu_i | \sum\lambda_i=1, \lambda_i\geq 0, i=0,1,...,k\right\},
\end{equation}
where $\{u_0,u_1,...,u_k\}\subset\mathbb{R}^d$ is a set of affinely independent points. Examples of $k$-simplex for the first few dimensions are shown in Figure \ref{fig:ph}. Essentially, a $0$-simplex is a vertex, a $1$-simplex is an edge, a $2$-simplex is a triangle, and a $3$-simplex is a tetrahedron. 
A {\it face} $\tau$ of $\sigma$ is the convex hull of a non-empty subset of $u_i$ and is {\it proper} if the subset does not contain all $k+1$ points. Equivalently, we can write as $\tau \le \sigma$ if $\tau$ is a face or $\sigma$, or $\tau < \sigma$ if $\tau$ is proper. The {\it boundary} of $\sigma$, is defined to be the union of all proper faces of $\sigma$.

\noindent
\paragraph{Simplicial complex} A {\it simplicial complex} is a finite collection of simplices $K$ such that $\sigma \in K$ and $ \tau \le \sigma$ implies $\tau \in K$, and $\sigma$, $\sigma_0 \in K$ implies $\sigma \cap \sigma_0$ is either empty or a face of both. The {\it dimension} of $K$ is defined to be the maximum dimension of its simplices.

\paragraph{Chain complex} Given a simplicial complex $K$ and a constant $p$ as dimension, a {\it $p$-chain } is a formal sum of $p$-simplices in $K$, denoted as $c=a_i\sigma_i$. Here $\sigma_i$ are the $p$-simplices and the $a_i$ are the coefficients, mostly defined as 0 or 1 (module 2 coefficients) for computational considerations. Specifically, $p$-chains can be added as polynomials. If $c_0=\sum a_i\sigma_i$ and $c_1=\sum b_i\sigma_i$, then $c_0+c_1=\sum(a_i+b_i)\sigma_i$, where the coefficients follow $\mathbb{Z}_2$ addition rules. The $p$-chains with the previous defined addition form an Abelian group and can be written as $(C_p, +)$. A {\it boundary operator} of a $p$-simplex $\sigma$ is defined as
\begin{equation}
\partial_p\sigma=\sum_{j=0}^{p}(-1)^j[ u_0,u_1,...,\widehat{u_j},...,u_p],
\end{equation}
where $[ u_0,u_1,...,\widehat{u_j},...,u_p]$ means that vertex $u_j$ is excluded in computation. Given a $p$-chain $c=a_i\sigma_i$, we have $\partial_p c=\sum a_i\partial_p\sigma_i$. Notice that $\partial_p$ maps $p$-chain to $\{p-1\}$-chain and that boundary operation commutes with addition, a {\it boundary homomorphism} $\partial_p: \sigma_p \to \sigma_{p-1}$ can be defined. The chain complex can be further defined using such boundary homomorphism as following:
\begin{equation}
\cdot\cdot\cdot \xrightarrow{\makebox[.27in]{}}C_{p+1} \xrightarrow{\makebox[.27in]{$\partial_{p+1}$}}C_{p}\xrightarrow{\makebox[.27in]{$\partial_{p}$}}C_{p-1} \xrightarrow{\makebox[.27in]{$\partial_{p-1}$}}\cdots\xrightarrow{\makebox[.27in]{$\partial_1$}}C_0\xrightarrow{\makebox[.27in]{$\partial_0$}}0.\\
\end{equation} 

\paragraph{Cycles and boundaries} A $p$-cycle is defined to be a $p$-chain $c$ with empty boundary ($\partial_p c=0$), and the group of $p$-cycles of $K$ is denoted as $Z_p = Z_p(K)$. In other words, $Z_p$ in the kernel of the $p$-th boundary homomorphism, $Z_p = \ker \partial_p$. A $p$-boundary is a $p$-chain, say $c$, such that there exists $d \in C_{p+1}$ and $\partial_p d =c$, and the group of $p$-boundaries is written as $B_p = B_p(K)$. Similarly, we can rewrite $B_p$ as $B_p = \mathrm{im} \partial_{p+1}$ since the group of p-boundaries is the image of the $(p+1)$-st boundary homomorphism.  
\paragraph{Homology groups} The fundamental lemma of homology says that the composition operator $\partial_{p}\circ\partial_{p+1}$ is a zero map \cite{edelsbrunner:2000}. With this lemma, we conclude that $\mathrm{im} \partial{p+1}$ is a subgroup of $\ker \partial_p$. Then the {\it $p$-th homology group} of simplicial complex is defined as the $p$-th cycle group modulo the $p$-th boundary group, 
\begin{equation}
H_p = Z_p/B_p
\end{equation}
and the {\it $p$-th Betti number} is the rank of this group, $\beta_p = \mathrm{rank} H_p$. Geometrically, Betti numbers can be used to describe the connectivity of given simplicial complexes. Intuitively, $\beta_0$, $\beta_1$ and $\beta_2$ are numbers of connected components, tunnels, and cavities, respectively, for the first few Betti numbers.

\paragraph{Filtration and persistence} A filtration of a simplicial complex $K$ is a nested sequence of subcomplexes of $K$.
\begin{equation}
\varnothing = K_0 \subseteq K_1 \subseteq ... \subseteq K_n=K.
\label{eqn:filtration}
\end{equation}
For each $i \le j$, there exists an inclusion map from $K_i$ to $K_j$ and therefore an induced homomorphism $f_p^{i,j}: H_p(K_i) \to H_p(K_j)$ for each dimension $p$. The filtration defined in Equation (\ref{eqn:filtration}) thus corresponds to  a sequence of homology groups connected by homomorphisms. 
\begin{equation}
0 = H_p(K_0) \to H_p(K_1) \to \cdots \to H_p(K_n) = H_p(K)
\end{equation}
for each dimension $p$. The $p$-th persistent homology groups are defined as the images of the homomorphisms induced by inclusion, 
\begin{equation}
H^{i,j}_p = \mathrm{im}f^{i,j}_p
\end{equation}
where $0 \le i \le j \le n$. In other words, $H^{i,j}_p$ contains the homology classes of $K_i$ that are still alive at $K_j$ for given dimension $p$ and each pair $i,j$. We can reformulate the $p$-th persistent homology group as
\begin{equation}
H^{i,j}_p = Z_p(K_i)/\left(B_p(K_j)\cap Z_p(K_i)\right).
\end{equation}
The corresponding $p$-th persistent Betti numbers are the ranks of these groups, $\beta^{i,j}_p = \mathrm{rank}H^{i,j}_p$.  
The birth, death and persistence of a Betti number carry important chemical and/or biological  information, which is the basic of the present method.    

\subsubsection{Persistent homology for characterizing molecules} 
As introduced before, persistent homology indeed reveals long lasting properties of a given object and offers a practical method for computing topological features of a space. In the context of toxicity prediction, persistent homology captures the underlying invariants and features of small molecules directly from discrete point cloud data. A intuitive way to construct simplicial complex from point cloud data is to utilize Euclidean distance, or essentially to use so-called ``Vietoris-Rips complex''. Vietoris-Rips complex is defined to be a simplicial complex whose $k$-simplices correspond to unordered ($k+1$)-tuples of points which are pairwise within radius $\epsilon$. 

However, a particular radius $\epsilon$ is not sufficient since it is difficult to see if a hole is essential. Therefor, it is necessary to increase radius $\epsilon$ systematically, and see how the homology groups and Betti-numbers evolve.  The persistence \cite{edelsbrunner:2000, zomorodian:2005} of each Betti number over the filtration can be recorded in barcodes \cite{CZOG05,Ghrist:2008}. The persistence of topological invariants observed from barcodes offers an important characterization of molecular structures. For instance, given the 3D coordinates of a small molecule, a short-lived Betti-0 bar may be the consequence of a strong covalent bond while a long-lived Betti-0 bar can indicate a weak covalent bond. Similarly,  a long-lived Betti-1 bar may represent a chemical ring. 
Such observations motivate us to design persistent homology based topological descriptors. However, it is important to note that the filtration radius is not a chemical bond and topological connectivity is not a physical relationship. In other word, persistent homology offers a representation of molecules that is entirely different from classical theories of chemical and/or physical bonds. Nevertheless, such a representation is systematical and comprehensive and thus is able to unveil structure-function relationships when it is coupled with  advanced machine learning algorithms.

\begin{figure}[ht!]
\small
\centering
\begin{subfigure}{0.3\textwidth}
\centering
\includegraphics[width=0.8\textwidth]{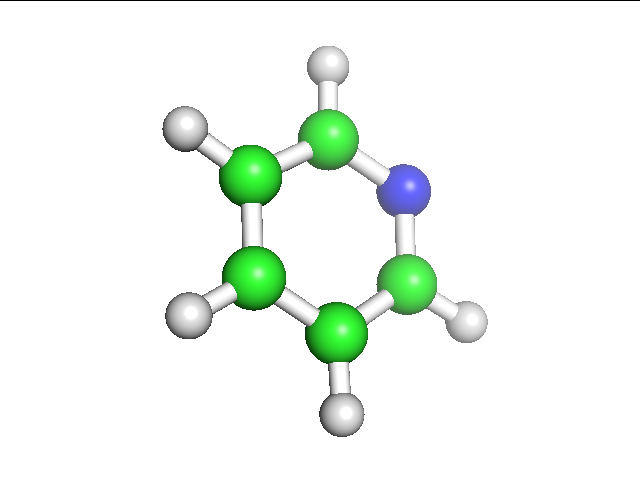}
\caption{All-elements}
\label{fig:pyridine}
\end{subfigure}
\centering
\begin{subfigure}{0.3\textwidth}
\centering
\includegraphics[width=0.8\textwidth]{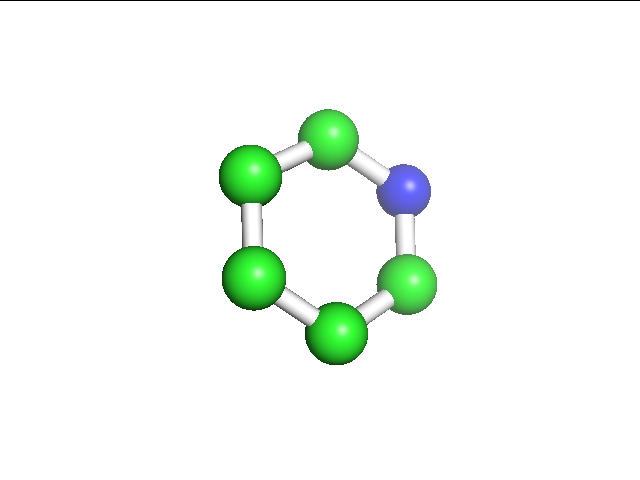}
\caption{C-N elements}
\label{fig:pyridine_noH}
\end{subfigure}
\begin{subfigure}{0.3\textwidth}
\centering
\includegraphics[width=0.8\textwidth]{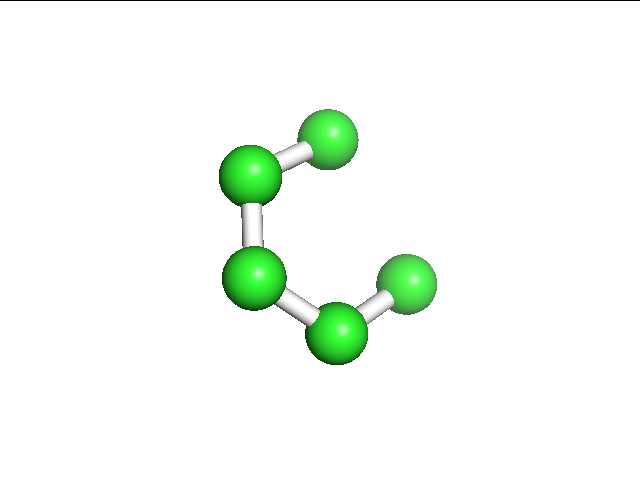}
\caption{C element}
\label{fig:pyridine_noN}
\end{subfigure}
\centering
\begin{subfigure}{0.3\textwidth}
\centering
\includegraphics[width=0.8\textwidth]{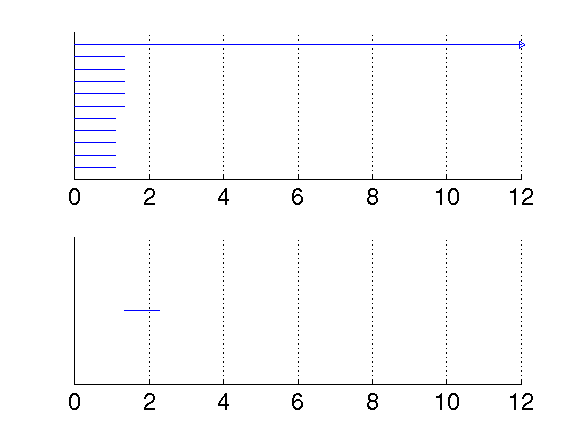}
\caption{Barcode plot of the all-elements}
\label{fig:bar-all}
\end{subfigure}
\centering
\begin{subfigure}{0.3\textwidth}
\centering
\includegraphics[width=0.8\textwidth]{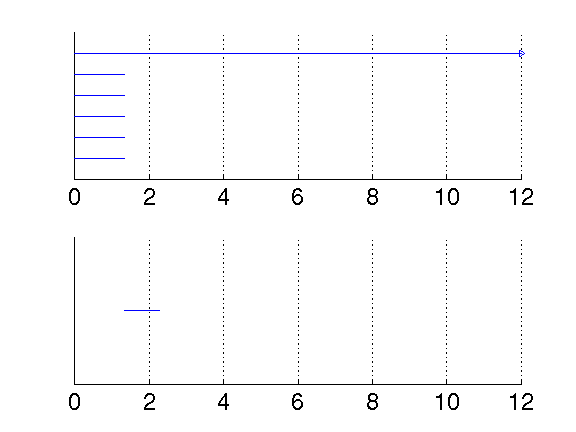}
\caption{Barcodes of the C-N elements}
\label{fig:bar-noH}
\end{subfigure}
\centering
\begin{subfigure}{0.33\textwidth}
\centering
\includegraphics[width=0.8\textwidth]{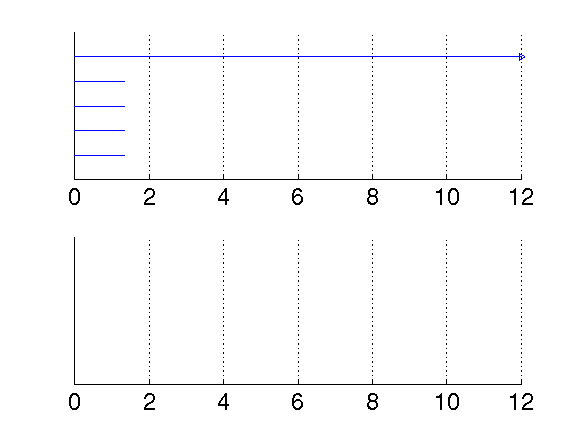}
\caption{Barcodes of the C element}
\label{fig:bar-cc}
\end{subfigure}
\caption{An illustration of pyridine and its persistent homology barcode plots.  In subfigure (a), (b) and (c), all atoms, carbon and nitrogen atoms, only carbon atoms are used for persistent homology computation, respectively. In subfigure (d), (e) and (f), from top to bottom, the results are computed for 0-dimension (Betti-0) and 1-dimension (Betti-1), respectively}
\label{fig:pyridine-example}
\end{figure}

\paragraph{An example} Figure \ref{fig:pyridine-example} is an detailed example of how our ESTDs are calculated and how they can reveal the structural information of pyridine. An all-elements representation of pyridine is given in Fig. \ref{fig:pyridine}, where carbon atoms are in green, nitrogen atom is in blue and hydrogen atoms are in white.  Without considering covalent bonds, there exist 11 isolated vertices (atoms) in Fig \ref{fig:pyridine}. Keep in mind that if the distance between two vertices is less than the filtration value then these two vertices do not connect. Thus at filtration value 0, we should have 11 independent components and no loops, which are respectively reflected by the 11 Betti-0 bars and 0 Betti-1 bars in Fig \ref{fig:bar-all}. As the filtration value increases to 1.08 \AA, every carbon atom starts to connect with its nearest hydrogen atoms, and consequently the number of independent components (also the number of Betti-0 bars) reduces to 6. When filtration value reaches 1.32 \AA, we are left with 1 Betti-0 bar and 1 Betti-1 bar. It indicates that there only exists one independent  component and the hexagonal carbon-nitrogen ring appears since the filtration value has exceeded the length of both carbon-carbon bond and carbon-nitrogen bond. As the filtration value becomes sufficiently large, the hexagonal ring is eventually filled and there is only one totally connected component left.  

It is worth to mention that Fig. \ref{fig:bar-all} does not inform the existence of the nitrogen atom in this molecule. Much chemical information is missing during the topological simplification. This problem becomes more serious as the molecular becomes larger and its composition becomes more complex. A solution to this problem is the element specific persistent homology or multicomponent persistent homology \cite{ZXCang:2017a}. In this approach, a molecule is decomposed into multiple components according to the selections of element types and persistent homology analysis is carried out on each component. The all-atom persistent homology shown in  Fig. \ref{fig:bar-all} is a special case in the multicomponent persistent homology. Additionally,  barcodes in Fig. \ref{fig:bar-noH} are for all carbon and nitrogen elements, while barcodes in  Fig. \ref{fig:bar-cc} are for carbon element only. By a comparison of these two barcodes, one can conclude that there is a nitrogen  atom in the molecule and it must be on the ring.  In this study, all persistent homology computations are carried out by Dionysus (\url{http://mrzv.org/software/dionysus/}) with Python bindings.

\paragraph{Element specific networks } The key to accurate prediction is to engineer ESTDs from corresponding element specific networks (ESNs) on which persistent homology is computed. As the example above shows, it is necessary to choose different element combinations in order to capture the properties of a given molecule. Carbon (C), nitrogen (N) and oxygen (O) are commonly occurring elements in small molecules.   Unlike proteins where hydrogen atoms are usually excluded  due to their absence in the database, for small molecules it is beneficial to include hydrogen atoms in  our ESTD calculations. Therefore ESNs of single-element types include four type elements ${\cal A}$=\{H,C,N,O\}. Additionally, we also consider   
 element combinations that involve two or more element types in an element specific network. In particular, the barcode of the network consisting of N and O elements in molecule might reveal  hydrogen bond interaction strength.  

\begin{table*}[!ht]
\centering
\caption{Element specific networks used to characterize molecules  }
\begin{tabular}{|c|c|}
\hline
Network type &  Element specific  networks \\ \hline 
Single-element  &          $\{a_i\}$, where $a_i \in {\cal A}$,  ${\cal A}$=\{H, C, N, O\}\\ \hline 
\multirow{2}{*}{Two-element}  & $\{b_i, c_j\}$, where $b_i \in {\cal B}$, $c_j \in {\cal C}$, $i \in \{1\ldots3\}$, $j \in \{1\ldots9\}$, and $i<j$. \\ 
& Here ${\cal B}$=\{C, N, O\} and ${\cal C}$=\{C, N, O, F, P, S, Cl, Br, I\}. \\ \hline
\end{tabular}
\label{tab:features}
\end{table*} 

Networks with a wide variety of element combinations  were tested and a good selection of such combinations is shown in Table \ref{tab:features}. Specifically,  two types of networks are used in the present work, namely, single-element networks and two-element networks.  Denote $a_i$ the $i$th atom of element type $a$ and $\{ a_i\}$ the set of all atoms of element type $a$ in a molecule.  Then   $\{ a_i\}$ with $a \in {\cal A}$ includes four different single-element type networks.  Similarly, Table \ref{tab:features} lists 21 different two-element networks. Therefore, a total of 25 element specific networks is used in the present work.

\paragraph{Filtration matrix} Another importance aspect  is the filtration matrix that defines the distance in persistent homology analysis  \cite{KLXia:2014c, ZXCang:2017d}.  We denote the Euclidean distance between atom $i$ at $(x_i, y_i, z_i)$ and atom $j$ at $(x_j, y_j, z_j)$ to be 
\begin{equation}
d_{i,j}= \sqrt{(x_i-x_j)^2+(y_i-y_j)^2+(z_i-z_j)^2}. 
\label{eqn:distance}
\end{equation}
By a direct filtration based on the Euclidean distance, one  can capture the information of covalent bonds easily as shown in  Fig. \ref{fig:bar-all}. 
However,  intramolecular interactions such as hydrogen bonds and van der Waals interactions cannot be revealed. In other words, the Betti-0 bar of two atoms with certain hydrogen bonding effect cannot be captured since there already exist shorter Betti-0 bars (covalent bonds). To circumvent such deficiencies we use filtration matrix to redefine the distance  
\begin{equation}
    M_{i, j} = 
	\begin{cases}
    		d_{i,j}, & \text{if \hspace{0.1in}} d_{i,j} \ge r_i + r_j +|\Delta d |\\
    		d_\infty,              & \text{otherwise},
	\end{cases}
	\label{eqn:newdistance}
\end{equation}
 where $r_i$ and $r_j$ are the atomic radius of atoms $i$ and $j$, respectively. Here $\Delta d$ is the bond length deviation in the data set and    $d_\infty$ is a large number which is set to be greater than the maximal filtration value.
By setting the distance between two atoms that have a covalent bond to a sufficiently large number, we are able to use topology to capture important  intramolecular interactions, such as hydrogen bonds, electrostatic interactions and van der Waals interactions.

\paragraph{Topological dimension} Finally we need to consider the dimensions of topological invariants. For large molecules such as proteins, it is important to compute the persistent homology of first three dimensions, which will result in Betti-0, Betti-1 and Betti-2 bars. The underlying reason is that proteins generally consists of thousands  of  atoms, and Betti-1 and Betti-2 bars usually contain very rich geometric information such as internal loops and cavities. However,  small molecules are geometrically  relatively simple and their  barcodes of high dimensions are usually very sparse. Additionally, small molecules are chemically complex due to their involvement of many element types and oxidation states. As such, high dimensional barcodes  of element specific networks carry little information. Therefore, we only consider Betti-0 bars for small molecule modeling.

\subsubsection{ESTDs for small molecules}

A general process for our ESTD calculation can be summarized as follows.
\begin{enumerate}
\item 3D coordinates of atoms of selected atom types are selected, and their Vietoris-Rips complexes are constructed. Note that distance defined in Eq. (\ref{eqn:newdistance}) is used for persistent homology barcodes generation.

\item The maximum filtration size is set to 10 \AA {\hspace{0.01in}} considering the size of small molecules. After  barcodes are obtained,  the first 10 small intervals of length 0.5 \AA {\hspace{0.01in}} are considered. In other words,   ESTDs will be calculated based on the barcodes of each subinterval $\mathrm{Int}_i = [0.5\mathrm{i}, 0.5(\mathrm{i}+1)]$, $i=0,\ldots,9$.

\begin{itemize}
\item Within each $\mathrm{Int}_i$, search Betti-0 bars whose birth time falls within this interval and  Betti-0 bars that dies within $\mathrm{Int}_i$, respectively and denote these two sets of Betti-0 bars as $\mathrm{S}_{\mathrm{birth}_i}$ and $\mathrm{S}_{\mathrm{death}_i}$.
\item Count the number of Betti-0 bars within $\mathrm{S}_{\mathrm{birth}_i}$ and $\mathrm{S}_{\mathrm{death}_i}$, and these two counts yield 2 ESTDs for the interval $\mathrm{Int}_i$.
\end{itemize}
\item In addition to interval-wise descriptors, we also consider global ESTDs for the entire barcodes. All Betti-0 bars' birth times and death times are collected and added into $\mathrm{S}_{\mathrm{birth}}$ and $\mathrm{S}_{\mathrm{death}}$, respectively. The maximum, minimum, mean and sum of each set of values are then computed as ESTDs. This step gives 8 more ESTDs.
\end{enumerate} 
Therefore for each element specific network, we have a total of 28 (2 $\times$ 10 intervals + 8) ESTDs. Since we consider a total 25 single-element and two-element networks, we have a total 700 (25 $\times$ 28) ESTDs.

 Finally, we would like to emphasize the essential ideas of our choice of ESTDs. In Step 2 of the ESTD generation process, we collect all birth and death time of Betti-0 bars in order to capture the hydrogen bonding and van der Waals interactions. These intramolecular interactions are captured by eliminating the topological connectivity of covalent bonds. The birth position can signal the formation of hydrogen bonding, and the death position represents the disappearance of such effects, which in turn reflects the strength of these effects. In step 3 of the above process, we consider all potential element-specific intramolecular effects together and use statistics of these effects as global descriptors for a given molecule. This would help us to better characterize small molecules. \par 
The topological feature vector that consists of ESTDs for the $i$-th molecule in the $t$-th prediction task (one task for each toxicity prediction), denoted as $\mathbf{x_i^t}$, can be used to approximate of the topological functional $f^t$ of MT-DNN. This optimization process will be carefully discussed in Section \ref{sec:MTL}.

\subsection{Auxiliary molecular descriptors} \label{sec:feature}

In addition to ESTDs, we are also interested in constructing a set of  microscopic features based on physical models to describe  molecular toxicity. This set of features should be  convenient for being used in different machine learning approaches, including deep learning and non deep learning, and single-task and multi-task ones. To make our feature generation feasible and robust to all compounds, we consider three types of basic physical information, i.e., atomic charges computed from quantum mechanics or molecular force fields, atomic surface areas calculated for solvent excluded surface definition, and atomic electrostatic solvation free energies estimated from the Poisson model.  To obtain this information, we first construct optimized 3D structure of for each molecule. Then the aforementioned atomic properties are computed. Our feature generation process can be divided into several steps:
\begin{enumerate}
\item {\bf Structure} Optimized 3D structures were prepared by LigPrep in \href{https://www.schrodinger.com/}{ Schr\"{ o}dinger suites (2014-2)}  from the original 2D structures, using options: \{-i 0 -nt -s 10 -bff 10\}. 

\item {\bf Charge} Optimized 3D structures were then fed in antechamber  \cite{wangjm:2006}, using parametrization: AM1-BCC charge, Amber mbondi2 radii and general Amber force field (GAFF) \cite{wangjm:2004}. This step leads to pqr files with corresponding charge assignments.

\item  {\bf Surface} \href{http://weilab.math.msu.edu/ESES/}{ESES online server}  \cite{ESES:2017}  was used to compute atomic surface area of each molecule, using pqr files from the previous step. This step also results in molecular solvent excluded surface information.  

\item {\bf Energy} \href{http://weilab.math.msu.edu/MIBPB/}{MIBPB online server}  \cite{DuanChen:2011a} was used to calculate the atomic electrostatic solvation free  energy  of each molecule, using surface and pqr files from previous steps. 
\end{enumerate}  
Auxiliary molecular descriptors were obtained according to the above procedure. Specifically, these molecular descriptors come from Step 2, Step 3 and Step 4. To make our method scalable and applicable to all kinds of molecules,  we manually construct element-specific molecular descriptors so that it does not depend on atomic positions or the number of atoms. The essential idea of such construction is to derive atomic properties of the each element type,  which is very similar to the idea of ESPH.

We consider 10 different commonly occurring element types, i.e., {\rm H, C, N, O, F, P, S, Cl, Br,} and {\rm I} and three different types of descriptors -- charge, surface area and electrostatic solvation free energies. Given an element type and a descriptor type, we compute the statistics of the quantities obtained from the aforementioned physical model calculation, i.e., summation, maximum, minimum, mean and variance, giving rise to 5 physical descriptors. To capture absolute strengths of each element descriptor, we further generate 5 more physical descriptors after taking absolute values of the same quantities. Consequently, we have a total of 10 physical descriptors for each given element type and descriptor type. Thus 300 (10 descriptor $\times$ 10 element types $\times$ 3 descriptor type) molecular descriptors can be generated at element type level.  

Additionally when all atoms are included for computation, 10 more physical descriptors can be constructed in a similar way (5 statistical quantities of original values, and another 5 for absolute values) for each element descriptor type (charge, surface area and electrostatic solvation free energies). This step yields another 30  molecular descriptors. As a result, we organize all of the above information into a 1D feature vector with 330 components, which is readily suitable for ensemble methods and DNN.  

These auxiliary molecular descriptors result in an independent descriptor set. When adding these molecular descriptors to the previously-mentioned ESTDs, we  have  a full descriptor set.

\subsection{Descriptor selection } \label{sec:fs}

The aforementioned descriptor construction process results in a large amount of descriptors, which naturally leads to the concern of descriptor ranking and overfitting. Therefore we rank all descriptors  according to their feature importance and use various feature importance thresholds as a selection protocol. Here the feature importance is defined to be Gini importance \cite{breiman:2001}  weighted by the number of trees in a forest calculated by our baseline method GBDT with scikit-learn \cite{scikit-learn}, and train separate models to examine their predictive performances on test sets. Four different values are chosen (2.5e-4, 5e-4, 7.5e-4 and 1e-4) and detailed analysis of their performances are also presented in a later section.

\subsection{Topological learning algorithms}
In this subsection, we integrate topology and machine learning to construct topological learning algorithms. Two types of  machine learning algorithms, i.e.,  ensemble methods and DNN algorithms are used in this study. Training details are also provided. 

\subsubsection{Ensemble methods} 

To explore strengths and weaknesses of different machine learning methods, we consider two popular ensemble methods, namely,  random forest (RF) and gradient boosting decision tree (GBDT). These  approaches have been widely used in solving QSAR prediction problems, as well as solvation and protein-ligand binding free energy predictions \cite{BaoWang:2016FFTS,BaoWang:2016FFTB,ZXCang:2017b}. They naturally  handle correlation between descriptors, and usually do not require a sophisticated feature selection procedure. Most importantly, both RF and GBDT are essentially  insensitive to parameters and robust to redundant features. Therefore,  we choose these two machine learning methods as baselines in our comparison. 

We have implemented RF and GBDT using the scikit-learn package (version 0.13.1)  \cite{scikit-learn}. The number of estimators is set to $2000$ and the learning rate is optimized for GBDT method. For each set, 50 runs (with different random states) were done and the average result is reported in this work.  Various descriptors groups discussed in Section \ref{sec:feature} are used as input data for RF and GBDT. More specifically, the maximum feature number is set to the square-root of the given descriptor length for both RF and GBDT models to facilitate training process given the large number of features, and it is shown that the performance of the average of sufficient runs is very decent. 

\subsubsection{Single-task deep learning algorithms  } \label{sec:DNN_arch}
A neural network acts as a transformation  that maps an input feature vector to an output vector. It essentially models the way a biological brain solves problems with numerous neuron units connected by axons. A typical shallow neural network consists of a few layers with neurons and uses back pr赵梦玥opogation to update weights on each layer. However, it is not able to construct hierarchical features and thus falls short in revealing more abstract properties, which makes it difficult to model complex non linear relationships.  

\begin{figure*}[!ht]
\small
\centering
\includegraphics[scale=0.6]{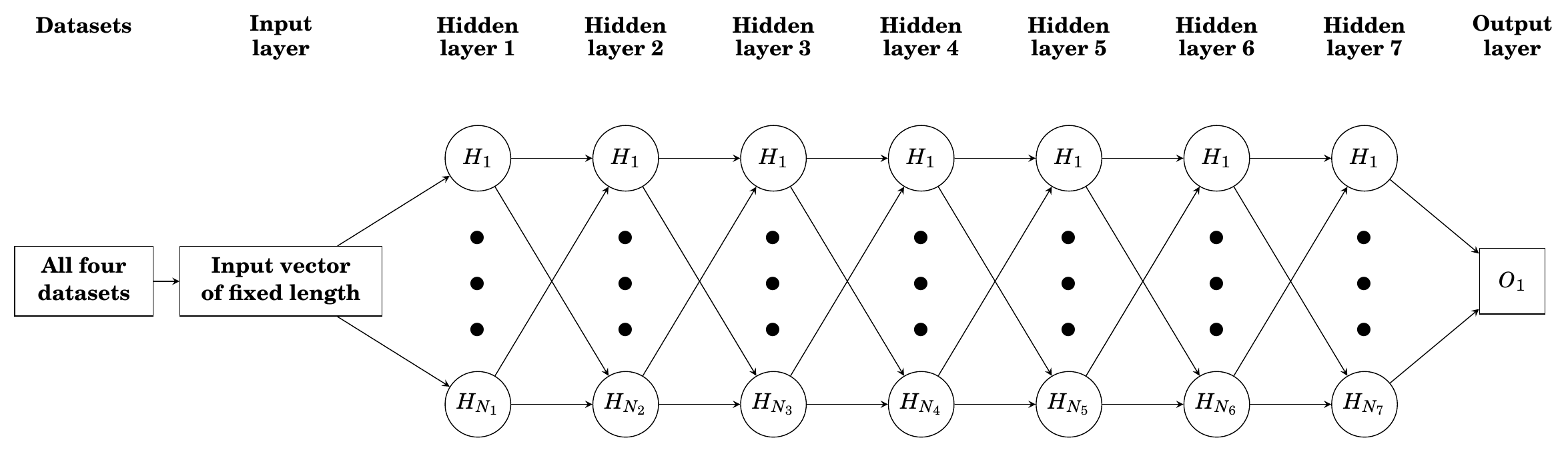}
\caption{An illustration of the ST-DNN architecture.}
\label{fig:st-DNN_arch}
\end{figure*}

A single-task deep learning algorithm, compared to shallow networks, has a wider and deeper architecture --  it consists of more layers and more neurons in each layer and reveals the facets of input features at different levels. Single-task deep learning algorithm is defined for each individual prediction task and only learns data from the specific task. A representation of such single task deep neural network (ST-DNN) can be found in Figure \ref{fig:st-DNN_arch}, where $N_i$ $(i=1,2,3)$ represents the number of neurons on the $i$-th hidden layer. 

\subsubsection{Multi-task learning} \label{sec:MTL}
Multi-task learning  is a machine learning technique which has shown success in qualitative Merck and Tox21 prediction challenges. The main advantage of MT learning is to learn multiple tasks simultaneously and exploit commonalities as well as differences across different tasks.  Another advantage of MT learning is that a small data set with incomplete statistical distribution to establish an accurate predictive model can often be significantly benefited from  relatively large data sets with more complete statistical distributions.

Suppose we have a total of $T$ tasks and the training data for the $t$-th task are denoted as $(\mathbf{x}_i^t, y_i^t)_{i=1}^{N_t}$, where $t={1,..,T}$, $i={1,...,N_t}$, $N_t$ is the number of samples of the $t$-th tasks, with $\mathbf{x}_i^t$ and $y_i^t$ being the  topological feature vector that consists of ESTDs and target  toxicity endpoint of the $i$-th molecule in $t$-th task, respectively.  The goal of MTL and topological learning is to minimize the following loss function for all tasks simultaneously:
\begin{equation}
\mathrm{argmin} \sum_{i=1}^{N_t} L(y_i^t, f^t(\mathbf{x}_i^t; \{\mathbf{W}^t,\mathbf{b}^t\} )) 
\end{equation}
where $f^t$ is a functional of the topological feature vector $\mathbf{x_i^t}$  parametrized by a weight vector $\mathbf{W}^t$ and bias term $\mathbf{b}^t$, and $L$ is the loss function. A typical cost function for regression is the mean squared error, thus the loss of the $t$-th task can be defined as:
\begin{align}\label{loss}
\mathrm{Loss \hspace{.03in} of \hspace{.03in}}  \mathrm{Task\hspace{.03in} }t &= \frac{1}{2}\sum_{i=1}^{N_t} L(\mathbf{x}_i^t, y_i^t) 
 = \frac{1}{2}\sum_{i=1}^{N_t} (y_i^t - f^t(\mathbf{x}_i^t; \{\mathbf{W}^t,\mathbf{b}^t\} )^2 
\end{align}
To avoid overfitting problem, it is usually beneficial to customize above loss function (\ref{loss}) by adding a regularization term on weight vectors, giving us an improved loss function for $t$-th task:
 \begin{align}\label{loss2}
\mathrm{Loss \hspace{.03in} of \hspace{.03in}}  \mathrm{Task\hspace{.03in} }t & = \frac{1}{2}\sum_{i=1}^{N_t} (y_i^t - f^t(\mathbf{x}_i^t; \{\mathbf{W}^t,\mathbf{b}^t\} )^2  + \beta||\mathbf{W}^t||_2^2
\end{align}
where $||\cdot||$ denotes the $L_2$ norm and $\beta$ represents a penalty constant. 

\begin{figure*}[!ht]
\small
\centering
\includegraphics[scale=0.6]{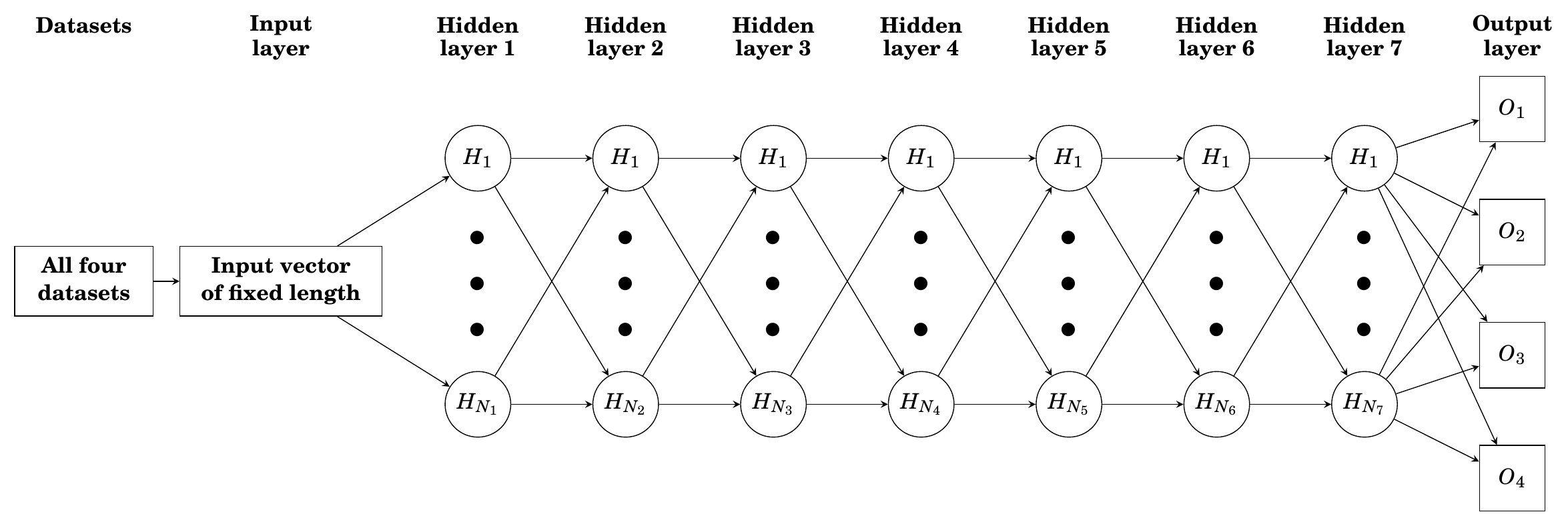}
\caption{An illustration of the MT-DNN architecture.}
\label{fig:mt-DNN_arch}
\end{figure*}

In this study, the goal of topology based MTL is to learn different toxicity endpoints jointly and potentially improve the overall performance of multiple toxicity endpoints prediction models. More concretely, it is reasonable to assume that different small molecules with different measured toxicity endpoints comprise distinct physical or chemical features, while descriptors such as the occurrence of certain chemical structure, can result in similar toxicity property. A simple representation of multitask deep neural network (MT-DNN) for our study is shown in Figure \ref{fig:mt-DNN_arch}, where $N_i$ $(i=1,\ldots, 7)$ represents the number of neurons on the $i$-th hidden layer, and $O_1$ to $O_4$ represent four predictor outputs. \par 
 
\subsubsection{Network parameters and training} \label{DNN_params}
Due to the large number of adjustable parameters, it is very time consuming to optimize all possible parameter combinations. Therefore we tune parameters within a reasonable range and subsequently evaluate their performances. The network parameters we use to train all models are: 4 deep layers with each layer having 1000 neurons, ADAM optimizer with 0.0001 as learning rate. It turns out that adding dropout or $L^2$ decay does not necessarily increase the accuracy and as a consequence we omit these two techniques. The underlying reason may be that the ensemble results of different DNN models is essentially capable of reducing bias from individual predictions.  A list of hyperparameters used to train all models can be found in Table \ref{tab:parameters}

\begin{table}[!ht]
\centering
\begin{tabular}{|c|c|}
\hline
Number of epochs & 1000 \\ \hline 
Number of hidden layers & 7 \\ \hline 
Number of neurons on each layer & 1000 for first 3 layers, and 100 for the next 4 layers \\ \hline 
Optimizer & ADAM \\ \hline  
Learning rate & 0.001 \\
\hline 
\end{tabular}
\caption{Proposed hyperparameters for MT-DNN}
\label{tab:parameters}
\end{table}

The hyperparameters selection of DNN is known to be very complicated. In order to come up with a reasonable set of hyperparameters, we perform a grid search of each hyperparameter within a wide range.  Hyperparameters in Table \ref{tab:parameters} are chosen so that we can have a reasonable training speed and accuracy.  In each training epoch,  molecules in each training set are randomly shuffled and then divided into mini-batches of size $200$, which are then used to update parameters. When all mini-batches are traversed, an training ``epoch" is done. All the training processes were done using Keras wrapper \cite{chollet:2015} with Theano (v0.8.2) \cite{theano} as the backend. All training were run on Nvidia Tesla K80 GPU and the approximate training time for a total of 1000 epochs is about 80 minutes.

\subsection{Evaluation criteria} 

Golbraikh {\it et al.}  \cite{golbraikh:2003} proposed a protocol to determine if a QSAR model has a predictive power. 
\begin{equation} \label{p0}
q^2 > 0.5,
\end{equation}
\begin{equation} \label{p1}
R^2 > 0.6, 
\end{equation}
\begin{equation} \label{p2}
\frac{R^2-R_0^2}{R^2} < 0.1 
\end{equation}
\begin{equation} \label{p3}
 0.85 \le k \le 1.15
\end{equation}
where $q^2$ is the  squared leave one out correlation coefficient for the training set, $R^2$ is the squared Pearson correlation coefficient between the experimental and predicted toxicities for the test set, $R_0^2$ is the squared  correlation coefficient between the experimental and predicted toxicities for the test set with the $y$-intercept being set to zero so that the regression is given by $Y=kX$. 
In addition to (\ref{p1}), (\ref{p2}) and (\ref{p3}), the prediction performance will also be evaluated in terms of root mean square error (RMSE) and mean absolute error (MAE). The prediction coverage, or fraction of chemical predicted, of corresponding methods is also taken into account since the prediction accuracy can  be increased by reducing the prediction coverage. 

\section{Results} \label{sec:results}

In this section, we first give a brief description of the data sets used in this work. We then carry out our predictions by using topological and physical features in conjugation with ST-DNN and MT-DNN,  and two ensemble methods, namely, RF and GBDT. The performances of these methods are compared with those of QSAR approaches used in the development of TEST software \cite{test_guide}. For the quantitative toxicity endpoints that we are particularly interested in, a variety of methodologies were tested and evaluated  \cite{test_guide}, including hierarchical method \cite{martin:2008}, FDA method, single model method, group contribution method \cite{martin:2001} and nearest neighbor method.   
 
As for ensemble models (RF and GBDT), the training procedure follows the traditional QSAR pipeline \cite{nantasenamat2009practical}. A particular model is then trained to predict the corresponding toxicity endpoint. Note that except for specifically mentioned, all our results shown in following tables are the average outputs of 50 numerical experiments. Similarly, to eliminate randomness in neural network training, we build 50 models for each set of parameters and then use their average output as our final prediction.  

Additionally, consensus of GBDT and MT-DNN is also calculated (the average of these two predictions) and its performance is also listed in tables for every dataset. Finally, the best results across all descriptor combinations are presented. 

\subsection{An overview of data sets}  \label{sec:data_construction}

This work concerns quantitative toxicity data sets. Four different quantitative toxicity  datasets, anmely, 96 hour fathead minnow LC$_{50}$ data set (LC$_{50}$ set), 48 hour Daphnia magna LC$_{50}$ data set (LC$_{50}$-DM set), 40 hour Tetrahymena pyriformis IGC$_{50}$ data set (IGC$_{50}$ set), and oral rat LD$_{50}$ data set (LD$_{50}$ set), are studied in this work. Among them, LC$_{50}$ set reports at  the concentration of test chemicals in water in mg/L that causes 50\% of fathead minnow to die after 96 hours. Similarly, LC$_{50}$-DM set records the concentration of test chemicals in water in mg/L that causes 50\% Daphnia maga to die after 48 hours. Both sets were originally downloadable from the ECOTOX aquatic toxicity database via web site \url{http://cfpub.epa.gov/ecotox/} and were preprocessed using filter criterion including media type, test location, etc  \cite{test_guide}.  The third set, IGC$_{50}$ set, measures the 50\% growth inhibitory concentration of Tetrahymena pyriformis organism after 40 hours. It was obtained from Schultz and coworkers \cite{akers:1999, zhuhao:2008}. The endpoint LD$_{50}$ represents the amount of chemicals that can kill half of rates when orally ingested. The LD$_{50}$ was constructed from ChemIDplus databse (\url{http://chem.sis.nlm.nih.gov/chemidplus/chemidheavy.jsp}) and then filtered according to several criteria \cite{test_guide}. 

The final sets used in this work are identical to those that were preprocessed and used to develop the  (\href{https://www.epa.gov/chemical-research/toxicity-estimation-software-tool-test}{Toxicity Estimation Software Tool} (TEST) \cite{test_guide}. TEST was developed to estimate chemical toxicity using various QSAR methodologies and is very convenient to use as it does not require any external programs. It follows the general QSAR workflow --- it first calculates 797 2D molecular descriptors and then predicts the toxicity of a given target by utilizing these precalculated molecular descriptors.  

\begin{table*}[!ht]
\centering
\caption{Statistics of quantitative toxicity data sets }
\begin{tabular}{|l|c|c|c|c|c|}
\hline
 & Total \# of mols &  Train set size & Test set size & Max value & Min value\\ \hline
LC$_{50}$ set & 823  & 659  & 164  & 9.261 & 0.037\\
LC$_{50}$-DM  set & 353 & 283 & 70  & 10.064 & 0.117  \\
IGC$_{50}$ set & 1792 & 1434 & 358  & 6.36 & 0.334 \\
LD$_{50}$ set & 7413 (7403) & 5931 (5924) & 1482 (1479)  & 7.201 & 0.291 \\ \hline
\end{tabular}
\label{train_result}
\end{table*}

All molecules are  in either 2D sdf format or SMILE string, and their corresponding toxicity endpoints are available on the TEST website. It should be noted that we are particularly interested in predicting quantitative toxicity endpoints so other data sets that contain qualitative endpoints or physical properties were not used. Moreover, different toxicity endpoints have different units. The units of LC$_{50}$, LC$_{50}$-DM, IGC$_{50}$ endpoints are $-\log_{10}({\mathrm{T \hspace{.02in} mol/L}})$, where {\rm T} represents corresponding endpoint. For LD$_{50}$ set, the units are $-\log_{10}({\mathrm{LD_{50} \hspace{.02in} mol/kg}})$. Although the units are not exactly the same, it should be pointed out that no additional attempt was made to rescale the values since endpoints are of the same magnitude order.  These four data sets also differ in their sizes, ranging from hundreds to thousands, which essentially challenges the robustness of our methods. A detailed statistics table of four datasets is presented in Table \ref{train_result}.

The number inside the parenthesis indicates the actual number of molecules that we use for developing models in this work. Note that for the first three datasets (i.e., LC$_{50}$, LC$_{50}$-DM and IGC$_{50}$ set), all molecules were properly included.  However, for LD$_{50}$ set, some molecules involved element {\rm As} were dropped out due to force field failure. Apparently, the TEST tool encounters  a similar problem  since results from two TEST models are unavailable, and the coverage (fraction of molecules predicted) from various TEST models is always smaller than one. The overall coverage of our models is always higher than  that of TEST models, which indicates a wider applicable domain of our models.

\begin{table*}[!ht]
\centering
\caption{Comparison of prediction results for the fathead minnow LC$_{50}$ test set.}
\begin{tabular}{|c|c|c|c|c|c|c|}
\hline
Method & $R^2$ & $\frac{R^2-R_0^2}{R^2}$ & $k$ & RMSE & MAE & Coverage \\ \hline
Hierarchical \cite{test_guide} & 0.710 & 0.075 & 0.966 & 0.801 & 0.574 & 0.951 \\ 
Single Model \cite{test_guide} & 0.704 & 0.134 & 0.960 & 0.803 & 0.605 & 0.945 \\ 
FDA  \cite{test_guide} & 0.626 & 0.113 & 0.985 & 0.915 & 0.656 & 0.945 \\ 
Group contribution \cite{test_guide} & 0.686 & 0.123 & 0.949 & 0.810 & 0.578 & 0.872 \\
Nearest neighbor  \cite{test_guide}& 0.667 & 0.080 & 1.001 & 0.876 & 0.649 & 0.939 \\
TEST consensus   \cite{test_guide} & 0.728 & 0.121 & 0.969 & 0.768 & 0.545 & 0.951 \\\hline
 \multicolumn{7}{|c|}{Results with ESTDs} \\ \hline
RF & 0.661 & 0.364 & 0.946 & 0.858 & 0.638 & 1.000 \\
GBDT  & 0.672 & 0.103 & 0.958 & 0.857 & 0.612 & 1.000 \\
ST-DNN & 0.675 & 0.031 & 0.995 & 0.862 & 0.601 & 1.000 \\
MT-DNN & 0.738 & 0.012 & 1.015 & 0.763 & 0.514 & 1.000 \\
Consensus & 0.740 & 0.087 & 0.956 & 0.755 & 0.518 & 1.000 \\ \hline 
 \multicolumn{7}{|c|}{Results with only auxiliary molecular descriptors} \\ \hline
RF & 0.744 & 0.467 & 0.947 & 0.784 & 0.560 & 1.000 \\
GBDT  & 0.750 & 0.148 & 0.962 & 0.736 & 0.511 & 1.000 \\
ST-DNN & 0.598 & 0.044 & 0.982 & 0.959 & 0.648 & 1.000 \\
MT-DNN & 0.771 & 0.003 & 1.010 & 0.705 & 0.472 & 1.000 \\
Consensus & 0.787 & 0.105 & 0.963 & 0.679 & 0.464 & 1.000 \\ \hline 
\multicolumn{7}{|c|}{Results with all descriptors} \\ \hline
RF & 0.727 & 0.322 & 0.948 & 0.782 & 0.564 & 1.000 \\
GBDT  & 0.761 & 0.102 & 0.959 & 0.719 & 0.496 & 1.000 \\
ST-DNN & 0.692 & 0.010 & 0.997 & 0.822 & 0.568 & 1.000 \\
MT-DNN & 0.769 & 0.009 & 1.014 & 0.716 & 0.466 & 1.000 \\ 
Consensus & {\bf 0.789} & 0.076 & 0.959 & {\bf 0.677} & {\bf 0.446} & 1.000 \\\hline 
\end{tabular}
\label{LC50_results}
\end{table*}

\subsection{Feathead minnow LC$_{50}$ test set}

The feathead minnow LC$_{50}$ set was randomly divided into a training set (80\% of the entire set) and a test set (20\% of the entire set) \cite{test_guide}, based on which a variety of TEST models were built. Table \ref{LC50_results} shows the performances of five TEST models, the TEST consensus obtained by the average of all independent TEST predictions,  four proposed methods and two consensus results obtained from averaging over present RF, GBDT, ST-DNN and MT-DNN results. TEST consensus gives the best prediction \cite{test_guide} among TEST results, reporting a correlation coefficient of 0.728 and RMSE of 0.768 log(mol/L). As Table \ref{LC50_results} indicates, our MT-DNN model outperforms TEST consensus both in terms of R$^2$ and RMSE with only ESTDs as input. When physical descriptors are independently used or combined with ESTDs, the prediction accuracy can be further improved to a higher level, with R$^2$ of 0.771 and RMSE of 0.705 log(mol/L). The best result is generated by consensus method using all descriptors, with R$^2$ of 0.789 and RMSE of 0.677 log(mol/L).

\subsection{Daphnia magna LC$_{50}$ test set} 
\begin{table*}[!ht]
\centering
\caption{Comparison of prediction results for the Daphnia magna LC$_{50}$ test set.}
\begin{tabular}{|c|c|c|c|c|c|c|}
\hline
 Method & $R^2$ & $\frac{R^2-R_0^2}{R^2}$ & $k$ & RMSE & MAE & Coverage \\ \hline
Hierarchical \cite{test_guide} & 0.695 & 0.151 & 0.981 & 0.979 & 0.757 & 0.886 \\ 
Single Model  \cite{test_guide}& 0.697 & 0.152 & 1.002 & 0.993 & 0.772 & 0.871 \\ 
FDA  \cite{test_guide}& 0.565 & 0.257 & 0.987 & 1.190 & 0.909 & 0.900 \\ 
Group contribution  \cite{test_guide}& 0.671 
& 0.049 & 0.999 & 0.803$^a$ & 0.620$^a$ & 0.657 \\
Nearest neighbor \cite{test_guide} & 0.733 & 0.014 & 1.015 & 0.975 & 0.745 & 0.871 \\
TEST consensus   \cite{test_guide}& 0.739 & 0.118 & 1.001 & 0.911 & 0.727 & 0.900 \\\hline
\multicolumn{7}{|c|}{Results with ESTDs} \\ \hline
RF & 0.441 & 1.177 & 0.957 & 1.300 & 0.995 & 1.000 \\
GBDT  & 0.467 & 0.440 & 0.972 & 1.311 & 0.957 & 1.000 \\
ST-DNN & 0.446 & 0.315 & 0.927 & 1.434 & 0.939 & 1.000 \\
MT-DNN & {\bf 0.788} & 0.008 & 1.002 & {\bf 0.805} & {\bf 0.592} & 1.000 \\
Consensus & 0.681 & 0.266 & 0.970 & 0.977 & 0.724 & 1.000 \\ \hline 
\multicolumn{7}{|c|}{Results with only auxiliary molecular descriptors} \\ \hline
RF & 0.479 & 1.568 & 0.963 & 1.261 & 0.946 & 1.000 \\
GBDT  & 0.495 & 0.613 & 0.959 & 1.238 & 0.926 & 1.000 \\
ST-DNN & 0.430 & 0.404 & 0.921 & 1.484 & 1.034 & 1.000 \\
MT-DNN & 0.705 & 0.009 & 1.031 & 0.944 & 0.610 & 1.000 \\
Consensus & 0.665 & 0.359 & 0.945 & 1.000 & 0.732 & 1.000 \\ \hline 
\multicolumn{7}{|c|}{Results with all descriptors} \\ \hline
RF & 0.460 & 1.244 & 0.955 & 1.274 & 0.958 & 1.000 \\
GBDT  & 0.505 & 0.448 & 0.961 & 1.235 & 0.905 & 1.000 \\
ST-DNN & 0.459 & 0.278 & 0.933 & 1.407 & 1.004 & 1.000 \\
MT-DNN & 0.726 & 0.003 & 1.017 & 0.905 & 0.590 & 1.000 \\
Consensus & 0.678 & 0.282 & 0.953 & 0.978 & 0.714 & 1.000 \\ \hline 
\end{tabular}\\
{$^a$ these values are inconsistent with $R^2=0.671$.}
\label{LC50DM_results}
\end{table*}

The Daphinia Magna LC50 set is the smallest in terms of set size, with 283 training molecules and 70 test molecules, respectively. However, it brings difficulties to building robust QSAR models given the relatively large number of descriptors. Indeed, five independent models in TEST software give significantly different predictions, as indicated by RMSEs shown in Table \ref{LC50DM_results} ranging from 0.810 to 1.190 log units. Though the RMSE of Group contribution is the smallest, its coverage is only  0.657 \% which largely restricts this method's applicability. Additionally, its $R^2$ value is inconsistent with its RMSE and MAE.  Since Ref. \cite{test_guide} states that ``The consensus method achieved the best results in terms of both prediction accuracy and coverage'', these usually low RMSE and MAE values might be typos. 

We also notice that our non-multitask models that contain ESTDs result in very large deviation from experimental values. Indeed, overfitting issue challenges traditional machine learning approaches especially when the number of samples is less than the number of descriptors. The advantage of MT-DNN model is to extract information from related tasks and our numerical results show that the predictions do benefit from MTL architecture. For models using ESTDs, physical descriptors and all descriptors, the R$^2$ has been improved from around 0.5 to 0.788, 0.705, and 0.726, respectively. It is worthy to mention that our ESTDs yield the best results, which proves the power of persistent homology. This result suggests that by learning related problems jointly and extracting shared information from different data sets, MT-DNN architecture can simultaneously perform multiple prediction tasks and enhances performances especially on small datasets.

\subsection{Tetraphymena pyriformis IGC$_{50}$ test set} 

IGC$_{50}$ set is the second largest QSAR toxicity set that we want to study. The diversity of molecules of in IGC$_{50}$ set is low and the coverage of TEST  methods is relatively high compared to   previous LC$_{50}$ sets. As shown in Table \ref{IGC50_results}, the $R^2$ of different TEST methods fluctuates from 0.600 to 0.764 and Test consensus prediction again yields the best result for TEST software with R$^2$ of 0.764. As for our models, the R$^2$ of MT-DNN with different descriptors spans a range of 0.038 (0.732 to 0.770), which indicates that our MT-DNN not only takes care of overfitting problem but also is insensitive to datasets. Although ESTDs slightly underperform compared to physical descriptors, its MT-DNN results are able to defeat most TEST methods except FDA method. When all descriptors are used, predictions by GBDT and MT-DNN outperform TEST consensus, with  R$^2$ of 0.787 and RMSE of 0.455 log(mol/L). The best result is again given by consensus method using all descriptors, with R$^2$ of 0.802 and RMSE of 0.438 log(mol/L).

\begin{table*}[!ht]
\centering
\caption{Comparison of  prediction results for the Tetraphymena Pyriformis IGC$_{50}$ test set.}
\begin{tabular}{|c|c|c|c|c|c|c|}
\hline
Method & $R^2$ & $\frac{R^2-R_0^2}{R^2}$ & $k$ & RMSE & MAE & Coverage \\ \hline
Hierarchical  \cite{test_guide}& 0.719 & 0.023 & 0.978 & 0.539 & 0.358 & 0.933 \\ 
FDA  \cite{test_guide}& 0.747 & 0.056 & 0.988 & 0.489 & 0.337 & 0.978 \\ 
Group contribution  \cite{test_guide}& 0.682 & 0.065 & 0.994 & 0.575 & 0.411 & 0.955 \\
Nearest neighbor  \cite{test_guide}& 0.600 & 0.170 & 0.976 & 0.638 & 0.451 & 0.986 \\
TEST consensus  \cite{test_guide} & 0.764 & 0.065 & 0.983 & 0.475 & 0.332 & 0.983 \\\hline
 \multicolumn{7}{|c|}{Results with ESTDs} \\ \hline
RF & 0.625 & 0.469 & 0.966 & 0.603 & 0.428 & 1.000 \\
GBDT  & 0.705 & 0.099 & 0.984 & 0.538 & 0.374 & 1.000 \\
ST-DNN & 0.708 & 0.011 & 1.000 & 0.537 & 0.374 & 1.000 \\
MT-DNN & 0.723 & 0.000 & 1.002 & 0.517 & 0.378 & 1.000 \\
Consensus & 0.745 & 0.121 & 0.980 & 0.496 & 0.356 & 1.000 \\ \hline 
 \multicolumn{7}{|c|}{Results with only auxiliary molecular descriptors} \\ \hline
RF & 0.738 & 0.301 & 0.978 & 0.514 & 0.375 & 1.000 \\
GBDT  & 0.780 & 0.065 & 0.992 & 0.462 & 0.323 & 1.000 \\
ST-DNN & 0.678 & 0.052 & 0.972 & 0.587 & 0.357 & 1.000 \\
MT-DNN & 0.745 & 0.002 & 0.995 & 0.498 & 0.348 & 1.000 \\
Consensus & 0.789 & 0.073 & 0.989 & 0.451 & 0.317 & 1.000 \\ \hline 
\multicolumn{7}{|c|}{Results with all descriptors} \\ \hline
RF & 0.736 & 0.235 & 0.981 & 0.510 & 0.368 & 1.000 \\
GBDT  & 0.787 & 0.054 & 0.993 & 0.455 & 0.316 & 1.000 \\
ST-DNN & 0.749 & 0.019 & 0.982 & 0.506 & 0.339 & 1.000 \\
MT-DNN & 0.770 & 0.000 & 1.001 & 0.472 & 0.331 & 1.000 \\ 
Consensus & {\bf 0.802} & 0.066 & 0.987 & {\bf 0.438} & {\bf 0.305} & 1.000 \\ \hline 
\end{tabular}
\label{IGC50_results}
\end{table*}

\subsection{Oral rat LD$_{50}$ test set} 

The oral rat LD$_50$ set contains the largest molecule pool with 7413 compounds. However, none of methods is able to provide a 100\% coverage of this data set. The results of single model method or group contribution method were not properly built for the entire set \cite{test_guide}.  It was noted that  LD$_{50}$ values of this data set are relatively difficult to predict as they have a higher experimental uncertainty \cite{zhuqsar:2009}. As shown in Table \ref{LD50_results}, results of two TEST approaches, i.e., Single Model and Group contribution, were not reported for this problem. The TEST consensus result improves overall prediction accuracy of other TEST methods by about 10 \%, however, other non-consensus methods all yield low R$^2$ and high RMSE. 

For our models, all results outperform those of non-consensus methods of TEST. In particular, GBDT and  MT-DNN with all descriptors yield the best (similar) results, giving slightly better results compared to TEST consensus. Meanwhile, our predictions are also relatively stable for this particular set as R$^2$s do not essentially fluctuate. It should also be noted that our ESTDs have slightly higher coverage than physical descriptors (all combined descriptors) since 2 molecules in the test set that contains As element cannot be properly optimized for energy computation. However this is not an issue with our persistent homology computation. Consensus method using all descriptors again yield the best results for all combinations, with optimal R$^2$ of 0.653 and RMSE of 0.568 log(mol/kg).

\begin{table*}[!ht]
\centering
\caption{Comparison of  prediction results for the Oral rat LD$_{50}$ test set.}
\begin{tabular}{|c|c|c|c|c|c|c|}
\hline
Method & $R^2$ & $\frac{R^2-R_0^2}{R^2}$ & $k$ & RMSE & MAE & Coverage \\ \hline
Hierarchical  \cite{test_guide}& 0.578 & 0.184 & 0.969 & 0.650 & 0.460 & 0.876 \\ 
FDA  \cite{test_guide}& 0.557 & 0.238 & 0.953 & 0.657 & 0.474 & 0.984 \\ 
Nearest neighbor \cite{test_guide} & 0.557 & 0.243 & 0.961 & 0.656 & 0.477 & 0.993 \\
TEST consensus  \cite{test_guide} & 0.626 & 0.235 & 0.959 & 0.594 & 0.431 & 0.984 \\\hline
\multicolumn{7}{|c|}{Results with ESTDs} \\ \hline
RF & 0.586 & 0.823 & 0.949 & 0.626 & 0.469 & 0.999\\
GBDT  & 0.598 & 0.407 & 0.960 & 0.613 & 0.455 &  0.999\\
ST-DNN & 0.601 & 0.006 & 0.991 & 0.612 & 0.446 & 0.999 \\
MT-DNN & 0.613 & 0.000 & 1.000 & 0.601 & 0.442 & 0.999 \\
Consensus & 0.631 & 0.384 & 0.956 & 0.586 & 0.432 & 0.999 \\ \hline 
\multicolumn{7}{|c|}{Results with only auxiliary molecular descriptors} \\ \hline
RF & 0.597 & 0.825 & 0.946 & 0.619 & 0.463 & 0.997\\
GBDT  & 0.605 & 0.385 & 0.958 & 0.606 & 0.455 & 0.997\\
ST-DNN & 0.593 & 0.008 & 0.992 & 0.618 & 0.447 & 0.997 \\
MT-DNN & 0.604 & 0.003 & 0.995 & 0.609 & 0.445  & 0.997\\
Consensus & 0.637 & 0.350 & 0.957 & 0.581 & 0.433 & 0.997 \\ \hline 
\multicolumn{7}{|c|}{Results with all descriptors} \\ \hline
RF & 0.619 & 0.728 & 0.949 & 0.603 & 0.452 &  0.997\\
GBDT  & 0.630 & 0.328 & 0.960 & 0.586 & 0.441 &  0.997\\
ST-DNN & 0.614 & 0.006 & 0.991 & 0.601 & 0.436 & 0.997\\
MT-DNN & 0.626 & 0.002 & 0.995 & 0.590 & 0.430  & 0.997\\ 
Consensus & {\bf 0.653} & 0.306 & 0.959 & {\bf 0.568} & {\bf 0.421} & 0.997 \\ \hline 
\end{tabular}
\label{LD50_results}
\end{table*}

\section{Discussion}
In this section, we will discuss how ESTDs bring new insights to quantitative toxicity endpoints and how ensemble based topological learning can improve overall performances. 

\subsection{The impact of descriptor selection and potential overfitting }

A major concern for the proposed models is descriptor redundancy and potential overfitting. To address this issue, four different sets of high-importance descriptors are selected  by a threshold to perform prediction tasks as described in Section \ref{sec:fs}. Table \ref{tab:LC50_fs} below shows the results of MT-DNN using these four different descriptor sets for LC50 set. Results for the other three remaining sets are provided in Supplementary information.
\begin{table*}[!ht]
\centering
\caption{Results of selected descriptor groups for LC50 set}
\begin{tabular}{|c|c|c|c|c|c|c|c|c|}
\hline
 Threshold & \# of descriptors & $R^2$ & $\frac{R^2-R_0^2}{R^2}$ & $k$ & RMSE & MAE & Coverage \\ \hline 
0.0 & 1030 &  0.769 & 0.009 & 1.014 & 0.716 & 0.466 & 1.000 \\
2.5e-4 & 411 &  0.784 & 0.051 & 0.971 & 0.685 & 0.459 & 1.000 \\
5e-4 &  308 & 0.764 & 0.062 & 0.962 & 0.719 & 0.470 & 1.000 \\
7.5e-4 & 254 &  0.772 & 0.064 & 0.958 & 0.708 & 0.468 & 1.000 \\
1e-3 & 222 & 0.764 & 0.063 & 0.963 & 0.717 & 0.467 & 1.000 \\ \hline 
\end{tabular}
\label{tab:LC50_fs}
\end{table*}

Table \ref{tab:LC50_fs} shows performance with respect to different numbers of descriptors. 
When the number of descriptors is increased from 222, 254, 308, 411 to 1030,  RMSE does not increase and $R^2$ does not change much.     
This behavior  suggests that our models are essentially insensitive to the number of descriptors and thus there is little overfitting. MT-DNN architecture takes care of overfitting issues by successive feature abstraction, which  naturally mitigates noise generated by less important descriptors. MT-DNN architecture can also potentially take advantage over related tasks, which in turn reduces the potential overfitting on single dataset by the alternative training procedure.  
 
Similar behaviors have also been observed for the remaining three datasets, as presented in Supplementary information. Therefore our MT-DNN architecture is very robust against feature selection  and can avoid  overfitting.

\subsection{The predictive power of ESTDs for toxicity}

One of the main objectives of this study is to understand toxicity of small molecules from a topological point of view. It is important to see if ESTDs alone can match those methods proposed in T.E.S.T. software. When all ESTDs (Group 6) and MT-DNN architecture are used for toxicity prediction, we observe following results:
\begin{itemize}
\item LC50 set and LC50DM set. Models using only ESTDs achieve higher accuracy than T.E.S.T. consensus method. 
\item LD50 set. Consensus result of ESTDs tops T.E.S.T. software in terms of both R$^2$ and RMSE and MT-DNN results outperform all non-consensus T.E.S.T methods.
\item IGC50 set. ESTDs are slightly underperformed than T.E.S.T consensus.  However, MT-DNN with ESTDs still yield better results than most non-consensus T.E.S.T methods except FDA.
\end{itemize}

It is evident that our ESTDs along with MT-DNN architecture have a  strong predictive power for all kinds of toxicity endpoints. The ability of MT-DNN to learn from related toxicity endpoints has resulted in a substantial improvement over ensemble methods such as GBDT. Along with physical descriptors calculated by our in-house MIBPB, we can obtain state-of-art results for all four quantitative toxicity endpoints.

\subsection{Alternative element specific networks for generating ESTDs}
Apart from the element specific networks proposed in Table \ref{tab:features}, we also use alternative element specific networks listed below in Table \ref{tab:new_features} to perform the same prediction tasks. Instead of using two types of element-specific networks, we only consider two-element networks to generate ESTDs, which essentially puts more emphasis on intramolecular interaction aspect. Eventually, this new construction yields 30 different element specific networks (9+8+7+6), and a total of 840 ESTDs (30 $\times$ 28) is calculated and used for prediction.  
\begin{table*}[!ht]
\centering
\caption{Alternative element specific networks used to characterize molecules}
\begin{tabular}{|c|c|}
\hline
Network type &  Element specific  networks \\ \hline 
\multirow{2}{*}{Two-element}  & $\{b_i, c_j\}$, where $b_i \in {\cal B}$, $c_j \in {\cal C}$, $i \in \{1\ldots3\}$, $j \in \{1\ldots9\}$, and $i<j$, \\ 
& where ${\cal B}$=\{H, C, N, O\} and ${\cal C}$=\{H, C, N, O, F, P, S, Cl, Br, I\}. \\ \hline
\end{tabular}
\label{tab:new_features}
\end{table*} 
On LC50 set, IGC50 set and LD50 set, overall performances of the new ESTDs can be improved slightly. However on LC50-DM set, the accuracy is comparably lower (still higher than T.E.S.T consensus). Detailed performances of these ESTDs are presented in Supplementary materials. Thus the predictive power of our ESTDs is not sensitive to the choice of element specific networks as long as reasonable element types are included.

\subsection{A potential improvement with consensus tools}

In this work, we also propose consensus method as discussed in Section \ref{sec:results}. The idea of consensus is to train different models on the same set of descriptors and average across all predicted values. The underlying mechanism is to take advantage of system errors generated by different machine learning approaches with a possibility to reduce bias for the final prediction. 

As we notice from Section \ref{sec:results}, consensus method offers a considerable boost in prediction accuracy. For reasonably large sets except LC50-DM set, consensus models turn out to give the best predictions. When it comes to small set (LC50-DM set), consensus models perform worse than MT-DNN. It is likely due to the fact that large number of descriptors may cause overfitting issues for most machine learning algorithms, and consequently generate large deviations, which eventually result in a large error of consensus method. Thus, it should be a good idea to preform prediction tasks with both MT-DNN and consensus methods, depending on the size of data sets, to take advantage of both approaches.

\section{Conclusion} \label{sec:conclusion}

Toxicity refers to the degree of damage a substance on an organism,  such as an animal, bacterium, or plant, and can be qualitatively or quantitatively measured by experiments. Experimental measurement of quantitative toxicity is extremely valuable, but is typically expensive and time consuming, in addition to potential ethic concerns. Theoretical prediction of quantitative toxicity has become a useful alternative in  pharmacology and environmental science. A wide variety of methods has been developed for toxicity prediction in the past. The performances of these methods depend  not only on the descriptors, but also on machine learning algorithms, which makes the model evaluation a difficult task.  

In this work, we introduce a novel method, called element specific topological descriptor (ESTD), for the characterization and prediction of small molecular quantitative toxicity. Additionally physical descriptors based on established  physical models are also developed  to enhance the predictive power of ESTDs. These new descriptors are integrated with a variety of advanced machine learning algorithms, including  two  deep neural networks (DNNs) and two ensemble  methods (i.e., random forest (RF) and gradient boosting decision tree (GBDT)) to construct topological learning strategies for   quantitative  toxicity analysis and prediction.  

Four quantitative toxicity data sets, i.e., 96 hour fathead minnow LC$_{50}$ data set (LC$_{50}$ set), 48 hour Daphnia magna LC$_{50}$ data set (LC$_{50}$-DM set), 40 hour Tetrahymena pyriformis IGC$_{50}$ data set (IGC$_{50}$ set), and oral rat LD$_{50}$ data set (LD$_{50}$ set), are used in the present study.  Comparison has also been made to the state-of-art approaches given in the literature \href{https://www.epa.gov/chemical-research/toxicity-estimation-software-tool-test }{Toxicity Estimation Software Tool} (TEST)\cite{test_guide}  listed by United States Environmental Protection Agency.  Our numerical experiments indicate that   the proposed ESTDs are as competitive as individual methods in T.E.S.T.  Aided with  physical descriptors and MT-DNN architecture, ESTDs  are able to establish state-of-art predictions for quantitative toxicity  data sets. Additionally,  MT deep learning algorithms are typically  more accurate than ensemble methods such as RF and GBDT. 

It is worthy to note that the proposed new descriptors are very easy to generate and thus  have almost 100\% coverage for all molecules, indicating their broader applicability to practical toxicity analysis and prediction. In fact, our topological descriptors are much easier to construct than physical descriptors, which depend on physical models and force fields. The present work indicates that ESTDs are a new   class of powerful descriptors for small molecules.

\section*{Availability}
Software for computing ESTDs and auxiliary molecular descriptors is available as online server at {\url{http://weilab.math.msu.edu/TopTox/} and {\url{http://weilab.math.msu.edu/MIBPB/}}, respectively. The source code for computing ESTDs can be found in  Supplementary materials. 

\section*{Supplementary materials} 
Detailed performances of four groups of descriptors based on feature importance threshold with MT-DNN are presented in Table S1-S4 of Supplementary materials. Results with the ESTDs proposed in the Section of Discussion  using different algorithms are listed in Table S5-S12 of Supplementary materials.

\vspace{1cm}
\section*{Funding information}

This work was supported in part by NSF Grants  IIS-1302285 and DMS-1721024  and MSU Center for Mathematical Molecular Biosciences Initiative.

\vspace{1cm}
\bibliography{refs}

\begin{center}
\begin{figure}
\centering
\includegraphics[scale=0.6]{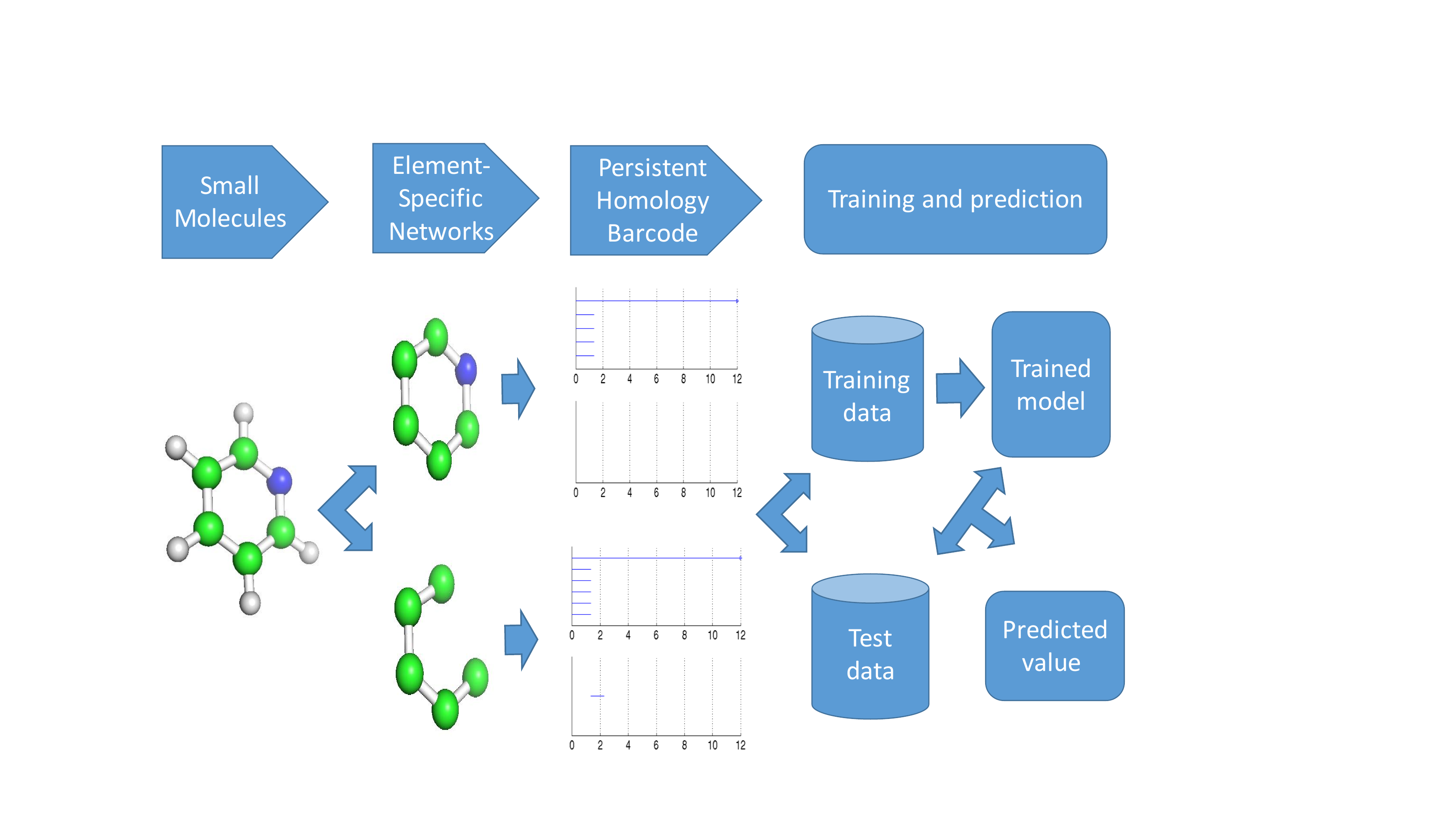}
\caption{Table of content graphic}
\end{figure}
\end{center}

\end{document}